\documentclass[conference,10pt]{IEEEtran}

\usepackage[noadjust]{cite}
\usepackage[pdftex]{graphicx}
\usepackage{amsmath}
\usepackage{algorithmic}
\usepackage{url}
\usepackage[bookmarks=false]{hyperref}
\usepackage{cleveref}
\usepackage{color}
\newcommand{\landauO}{\mathcal{O}}

\begin{document}

\title{A Flexible Network Approach to Privacy\\of Blockchain Transactions}

\author{\IEEEauthorblockN{David M\"odinger, Henning Kopp, Frank Kargl and Franz J. Hauck}
\IEEEauthorblockA{Institute of Distributed Systems, Ulm University\\
Email: \{david.moedinger, henning.kopp, frank.kargl, franz.hauck\}@uni-ulm.de}}

\maketitle

\begin{abstract}
For preserving privacy, blockchains can be equipped with dedicated mechanisms to
anonymize participants.
However, these mechanism often take only the abstraction layer of
blockchains into account whereas observations of the underlying network
traffic can reveal the originator of a transaction request.
Previous solutions either provide topological privacy that can be broken by
attackers controlling a large number of nodes, or offer strong and
cryptographic privacy but are inefficient up to practical unusability.
Further, there is no flexible way to trade privacy against efficiency to adjust
to practical needs.
We propose a novel approach that combines existing mechanisms to have
quantifiable and adjustable cryptographic privacy which is further improved by
augmented statistical measures that prevent frequent attacks with lower resources.
This approach achieves flexibility for privacy and efficency requirements
of different blockchain use cases.
\end{abstract}

\section{Introduction}

In recent years, more and more cryptocurrencies and other blockchain-based
technologies aim to provide privacy for their users, as the contents of
the blockchain may reveal sensitive information. Such information
could be purchasing behavior, credit balances, and how the cash has been
acquired~\cite{meiklejohn2013fistful,ron2013quantitative}. Some of these approaches use ring
signatures~\cite{saberhagen2013cryptonote,monerolab2014mysterious,noether2015ringct,kopp2017design}
or zero-knowledge proofs~\cite{miers2013zerocoin,ben2014zerocash} to achieve unlinkable
payments. Even already existing blockchains like Bitcoin are augmented with
privacy enhancing mechanisms~\cite{ruffing2014coinshuffle,bonneau2014mixcoin}.
However, these systems solely examine privacy by considering the blockchain and
its embedded transactions, leaving the underlying network vulnerable as it is
used to disseminate transactions within the peer-to-peer network of
nodes~\cite{koshy2014analysis}. Uncovering the IP address of the originator
of a transaction is a serious threat to user privacy as this address can be mapped
to real world identities.

Observing and analyzing network traffic within the peer-to-peer network of a
blockchain-based system can be done by injecting observer nodes into the
network. A small number of nodes with many interconnects or a larger number of
nodes, as they can be deployed by renting botnets, are a rather cheap way to
link a reasonably high percentage of the originators of submitted transactions
to IP addresses~\cite{Biryukov2014deanobitcoin}.
To prevent these attacks, further privacy mechanisms on the network
layer are required.

\begin{figure}[ht]
\vspace{1.5em}
\centering
\def\svgwidth{1\columnwidth}
\includegraphics[width=.9\columnwidth]{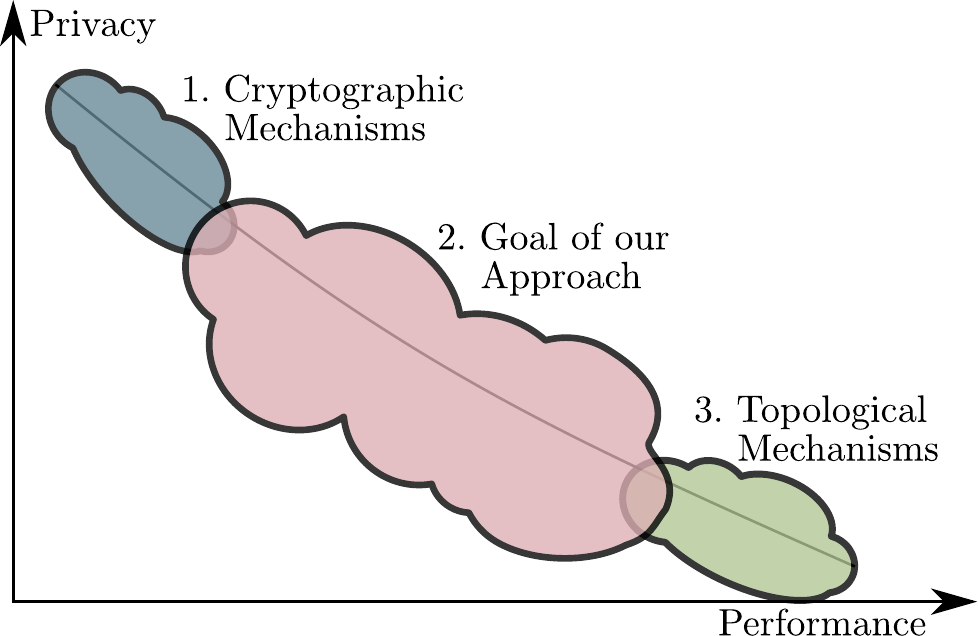}
\caption{The privacy-performance landscape. We are aware that privacy is dependent on attacker models and cannot be pin-pointed to a specific spot in a diagram. Thus, this diagram is just an illustration of our goals.}
\label{fig:overview}
\end{figure}

In the following \Cref{sec:scenario} we introduce the scenario and the context of this article.
In \Cref{sec:rec}, we outline some existing solutions and their drawbacks.
The considered systems can be divided into two main categories.
One category of systems provide inefficient but strong cryptographic
privacy. Their privacy guarantees are independent of the number of
observer nodes an attacker has under control (see 1. in Fig.~\ref{fig:overview}).
The second type of system achieves privacy by topological means, i.e.,
breaking the propagation symmetry of broadcasts. Approaches of this type are very
efficient, but can be defeated by an attacker controlling enough nodes
in the network (see 3. in Fig.~\ref{fig:overview}).

We provide our novel approach in \Cref{sec:approach}, which
combines these two types of systems
to achieve quantifiable and adjustable strong privacy (see 2. in Fig.~\ref{fig:overview}).
Thus, we utilize statistical measures to prevent cheap
and frequent attacks that can be executed, e.g., by using botnets.
With sophisticated attackers controlling or eavesdropping on large parts of the
network (e.g., intelligence agencies) our approach falls back to the
cryptographic privacy mechanisms which guarantee what is called k-anonymity. The strength of
this base privacy level and the associated cost depend on the size of
the parameter k, typically a value between four and ten.

In \Cref{sec:discussion}, we provide a brief discussion of stronger
attacker models and argument for the privacy and performance of our approach.
Lastly, \Cref{sec:conclusion} provides a short overview of our proposal and
an outlook for required work to progress towards a full solution.

\section{Scenario}
\label{sec:scenario}

Blockchains are the underlying technology originally introduced through
the digital cash system Bitcoin~\cite{nakamoto2009bitcoin}.
They implement a distributed append-only database, also called a ledger, with an
incorporated consensus mechanism which is used to agree on a global state.
In the case of Bitcoin the global state is the transaction history of tokens.
In systems like Ethereum~\cite{wood2014ethereum}, more general payloads,
such as the current state of a distributed state machine, are allowed.
We will refer to these payloads as transactions, though they may be
more general than financial transactions.
When a node wants to persist a transaction in the blockchain, it
broadcasts its transaction in a peer-to-peer network connecting all
participating nodes.
Some nodes in the network, called miners, verify the received
transactions, bundle them together with other transactions into blocks, and
vote by a procedure called proof of work for the inclusion of the block
into the blockchain.
If the block is included, the miner receives a financial reward for
having proposed the block, together with a small fee included in
each transaction. These transaction fees poses an incentive
to commit the transaction in a block, instead of generating empty
blocks.

In order to append blocks to the blockchain, miners need to have access to the
current global state, i.e., the latest block in the chain.
Thus it is very important that the broadcast of new blocks has a low
latency. This provides fairness to the miners, since otherwise miners
with high latency are disadvantaged in finding the next block and thus
collecting rewards.
Since all mechanisms to hide the originator of a block increase
latency and thus decrease fairness, we do not consider their privacy
in the remainder of this article.

On the level of transactions there is a similar trade-off.
Each transaction needs to be broadcast to all miners with low latency, such that each
miner has the same chance to earn the associated transaction fee.
Additionally, the user thereby decreases the time for his transaction
to be included in the blockchain.
In contrast to the transmission of blocks, the transmission of transactions calls
for stronger privacy, as they leak personal and
sensitive information~\cite{meiklejohn2013fistful,ron2013quantitative}.
While latency of the transaction propagation increases unfairness of
the earning of transaction fees, the reward for generating a block is
a lot higher. Therefore, latency is less of an issue.

Many blockchain applications apply methods to enhance transactions 
with privacy enhancing technologies, such as ring signatures
or zero-knowledge proofs.
Approaches to deanonymize transactions, e.g., link senders to IP addresses,
have made progress~\cite{Biryukov2014deanobitcoin,mastan2018deanobitcoin2}
rendering these privacy techniques on blockchain level incomplete.
Thus, we are in need of an anonymous broadcast mechanism for
transaction data in blockchain systems.

\section{Current Systems}
\label{sec:rec}
In this section we discuss existing systems for efficient topological
approaches to privacy for blockchain transactions on the network
layer, as well as cryptographic systems that can withstand stronger
attackers, but are less efficient.

The prominent anonymous communication system
Tor~\cite{dingledine2004tor} is usually one of the first approaches when trying
to achieve privacy on the network layer.
Tor only supports a direct connection between a pair of nodes
and does not provide an abstraction layer for broadcast communication.
While it would be possible to tunnel all connections through Tor,
it is not well suited to implement a broadcast mechanism for blockchains. Tor
can also be used in addition to the presented systems for a
defense in depth approach and will therefore not be discussed further in this paper.

\subsection{Topological Privacy Mechanisms}

\begin{figure}[htbp]
\vspace{1.5em}
\centering
\includegraphics[width=171.24230957bp]{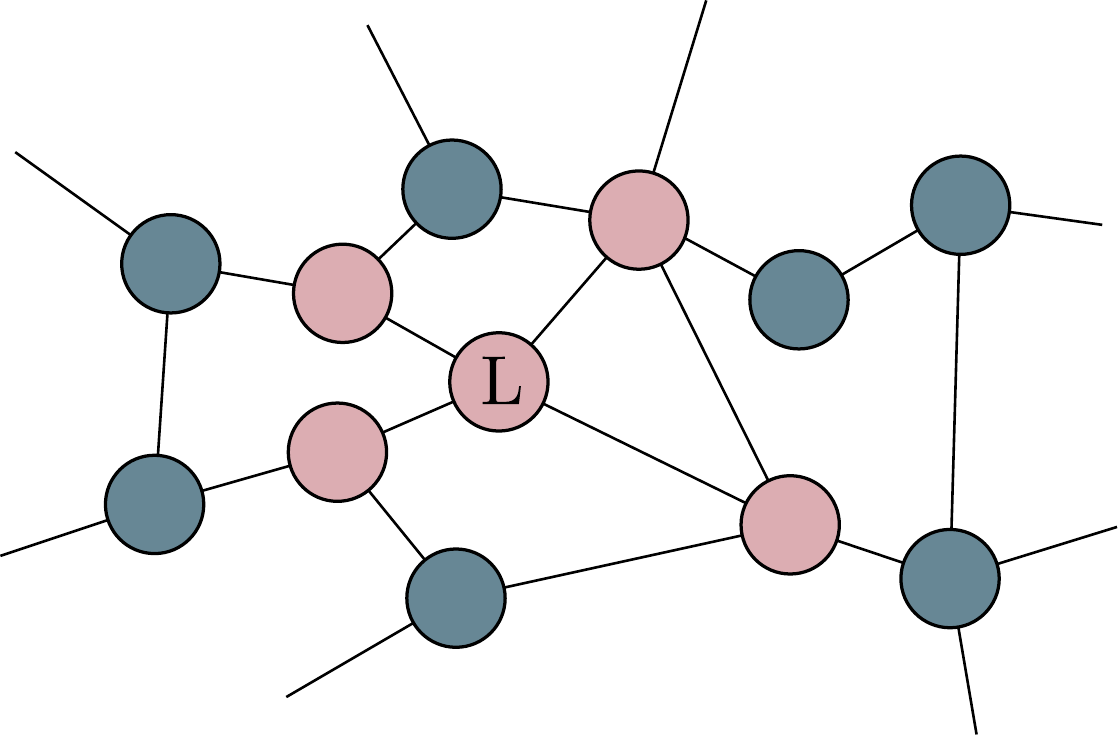}
\caption{A broadcast in progress. Light red nodes already received
the broadcast, while dark blue nodes did not. The likely originator is
marked with an L.}
\label{fig:scenario}
\end{figure}

Topological privacy mechanisms were introduced to prevent a
cheap and easy attack on anonymity, that can be employed using botnets.
These botnet-based attacks exploit the symmetry in propagation of
information dissemination when using broadcasts by observing
the network, e.g., by adding nodes until they control around \(20\%\)
of the network, and recording the arrival time of the received 
transactions~\cite{Biryukov2014deanobitcoin}. This attack works
even for so called unreachable\footnote{Nodes that refuse incoming connections.}
nodes~\cite{mastan2018deanobitcoin2}.
The result is a deanonymization of the transaction origin due to the
strongly skewed probabilities from the aforementioned propagation symmetry.
See Figure~\ref{fig:scenario} for an example.

One of the mechanisms to defend against these kind of attacks are topological privacy
mechanisms, sometimes also called statistical spreading mechanisms. This class
of mechanisms smoothes the likelihood of a node being the originator of a message
throughout the network. Instead of having one or few nodes in the center of the
graph of nodes that already received the message, all nodes are close to
equally likely the originator.

\textbf{Adaptive diffusion}~\cite{Fanti2015adaptivediffusion}, one of the
statistical spreading protocols, breaks the symmetry by creating a virtual
source token and spreading messages in such a way that the node currently owning
the virtual source token is the center of the spanned graph of nodes that
already received the message. The true source of the message can then be located
anywhere inside the graph. Adaptive diffusion keeps this graph balanced with the
node currently holding the virtual source token at the center after all steps.
The protocol consists of two alternating steps:
\begin{enumerate}
 \item Transfer the virtual source token with probability \(\alpha\) to a new node.
 \begin{itemize}
  \item \(\alpha\) is dependent on the number of rounds already executed for this message.
  \item After transferring the virtual source token, the new virtual source
  spreads the message in all directions besides the direction from which it received
  the virtual source token.
 \end{itemize}
 \item Spread the message further, increasing the diameter of the graph.
\end{enumerate}
The dissemination is accelerated by reducing \(\alpha\) after each round, as transmissions
of the virtual source token stalls the dissemination. By not spreading the message
via the previous virtual source, the graph of nodes having already received the message
has the new virtual source at its center. This is referred to as balancing.
These mechanisms smooth the probability of origin for every node.
Although this approach is designed for cycle-free networks,
measurements show it works well even for general networks~\cite{Fanti2015adaptivediffusion}.
One drawback, however, is that adaptive diffusion does not guarantee delivery of messages
to all nodes. In the context of blockchains these messages are transactions and failures
to deliver them to all nodes leads to unfairness as described in \Cref{sec:scenario}.

\begin{figure}[htbp]
\vspace{1.5em}
\centering
\includegraphics[width=171.24230957bp]{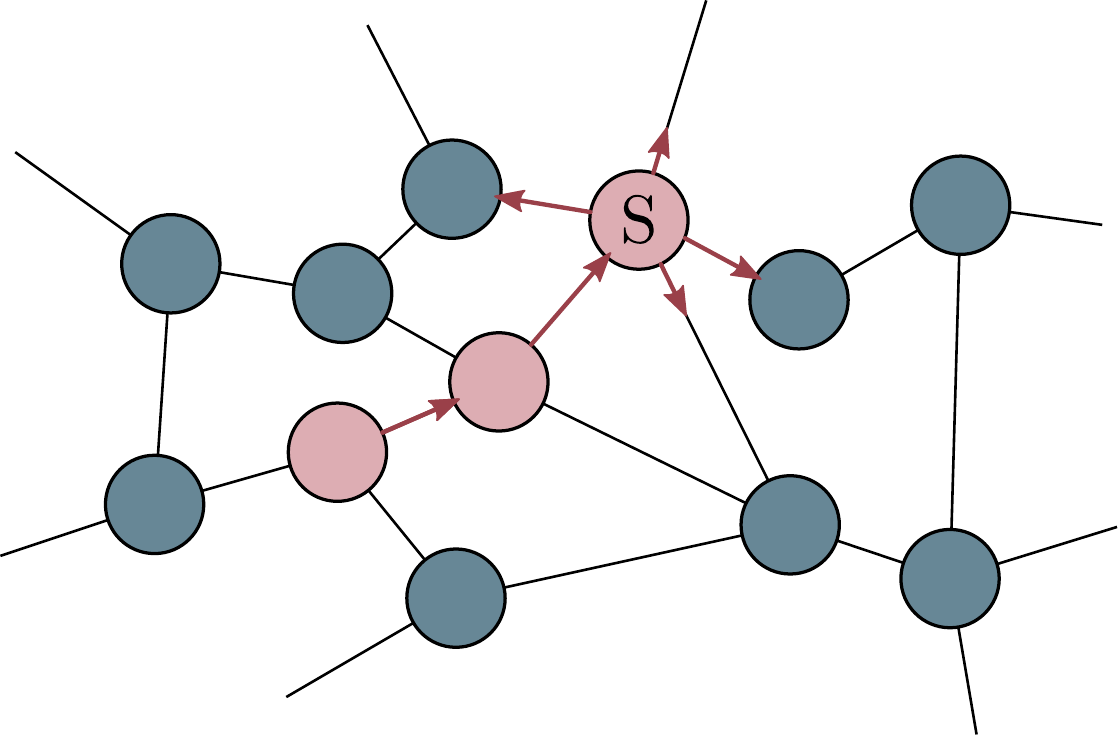}
\caption{An example of a Dandelion dissemination. The light red nodes have received the message along a line.
The last node S starts spreading the message in a regular broadcast manner.}
\label{fig:dandelion}
\end{figure}

With a goal of potential adoption by Bitcoin and guaranteed delivery,
\textbf{Dandelion}~\cite{Bojja2017dandelion} proposes a two phase
protocol for statistical spreading. Phase 1 spreads the transaction along a line graph. This
line is generated as an approximation of an Hamiltonian path. Phase 2 uses a regular 
flood and prune broadcast starting from the last node of the first phase.
\Cref{fig:dandelion} visualizes the switch from Phase 1 to Phase 2.
Anonymity is guaranteed through the first phase by transforming the
spreading graph into a linear path which is hard to observe and smooths
the probability of origin. Phase 2 ensures delivery to all nodes.
To protect against topology leaks in the first phase, which weaken the
anonymity properties, the creation of the Hamiltonian path approximation is repeated
periodically.

\subsection{Cryptographic Privacy Mechanisms}

Topological privacy mechanisms work well for smaller fractions of adversaries,
e.g., \(0.15\) to \(0.35\)~\cite{Bojja2017dandelion}, but provide little
privacy for large fractions of adversaries, even when there are known trustworthy
nodes left. To prevent against powerful attackers, there are some systems offering
privacy by means of cryptography, which can achieve privacy
independent of the number of observed links, and independent of the
computational power of the attacker.

A major building block used in these
systems~\cite{vonahn2003kanonmessages,Corrigan2010dissent,wolinsky2012dissent}
is the \textbf{dining cryptographers} network (DC-net)~\cite{Chaum1988dc}.
We provide one possible implementation of such an DC-net, which allows a fully
connected clique of nodes to anonymously broadcast one message of a bounded size
per round. A DC-net can be conceived as an operator that computes the bitwise XOR of
all input messages. Using this interpretation it is clear that only one message
can be sent per round, all other members need to set their message \(m=0\).

All nodes need to share pairwise encrypted channels.
The algorithm of \Cref{alg:dc} is computed by every member of the group.
\begin{figure}[h]
 \begin{algorithmic}
 \renewcommand{\algorithmicrequire}{\textbf{Input:}}
 \renewcommand{\algorithmicensure}{\textbf{Output:}}
 \REQUIRE Message \(m\), Group members \(G=\{g_1,g_2,\ldots,g_k\}\), maximum message length \(n\)
 \ENSURE Share \(m\) with all group members \(G\) or receive a message \(m\) sent by another group member.
 \STATE 1. Generate \(r_1,\ldots,r_k\) at random and of length \(n\), such that \(m=\bigoplus_{i=1\ldots k} r_i\).
 \STATE 2. Send \(r_i\) to \(g_i\) for \(i=1\ldots k\).
 \STATE 3. Collect the information from \(g_i\) as \(s_i\) for \(i=1\ldots k\).
 \STATE 4. Compute \(S=\bigoplus_{i=1..k}s_i\).
 \STATE 5. Send \(S\oplus s_i\) to \(g_i\) for \(i=1\ldots k\).
 \STATE 6. Collect the accumulation from \(g_i\) as \(t_i\) for \(i=1\ldots k\).
 \STATE 7. Compute \(T=\bigoplus_{i=1..k}t_i\).
 \STATE 8. Send \(T\oplus t_i\) to \(g_i\) for \(i=1\ldots k\).
 \STATE 9. Recover message \(m=T\oplus S\).
 \end{algorithmic}
 \caption{A possible implementation of a DC-network round of size \(k+1\), 
 executed by every group member separately.
 If a group member has no message to share, they use \(m=0\). If only one group member tried to
 send a message \(m\not=0\) it can be recovered as \(m=T\oplus S\).}
 \label{alg:dc}
\end{figure}

If \(T\oplus S\not=0\), someone sent a message. If there is only one sender, 
the message will be \(m=T\oplus S\). Potentially, multiple senders may have tried to send a message.
To detect this, message should carry CRC bits or a similar protection. On collision,
sending of messages has to be repeated with a backoff time.
The major drawbacks of DC-nets are that they do not scale well due to the quadratic
number of messages per participant, and their need for additional mechanisms to
prevent denial of service through malicious collisions. Von Ahn et al.~\cite{vonahn2003kanonmessages}
introduce a possibility to scale DC-nets by restricting it to k-anonymity and adding a
blame protocol, which detects misbehavior.

The state-of-the-art anonymity system
\textbf{Dissent}~\cite{Corrigan2010dissent,wolinsky2012dissent} provides
similar properties with a small amount of core servers as anonymity providers
and an anonymous announcement phase per round, where every participant announces
the length of message they want to transmit. This allows for variable sized
messages. The announcement phase uses a secure group shuffle for all: All nodes encrypt
their announcement with layers for all participants according to a fixed
permutation of the users per round. Each node then in turn shuffles all values and
removes their respective layer of encryption. After every node performed such a shuffle,
all nodes can trust the shuffle, since they participated, and only one
honest shuffle is necessary to hide the originator. The last participant publishes
the list of message lengths. With this information they perform a DC-round
to transmit the actual data.
The announcement round causes a startup phase~\cite{Corrigan2010dissent} scaling
linearly in the number of group members and becoming noticeably slow, e.g., 30 seconds,
for group sizes of 8 to 12. This latency might not be acceptable in real world
blockchain applications.

Networks for different applications, e.g., \textbf{Herd}~\cite{leblond2015herd}
for voice over IP or \textbf{RAC}~\cite{mokhtar2013dissentupdate},
use other building blocks, such as mix nodes, trust
zones, multiple rings with onion encryption and cover traffic. Whereas these could be used
for designing a privacy-preserving broadcast mechanism, they create
different problems. Mix nodes lead to increased load for central
infrastructure, due to their need to process all traffic. Cover
traffic creates continuous load, which is a problem for rare network utilization,
such as transaction transmissions and limits other uses as the rate is
dependent on the use case. While small DC-nets might scale the amount of rounds
depending on usage, scaling of cover traffic is much more restricted. If a
node increases or reduces their bandwidth consumtion, this change in behavior can be
attributed to their personal change in usage and is not reduced to a change in behavior
of members of a group.
This leaks the information of data usage and in the context of blockchain systems
can be correlated to the arrival times of new transactions, undermining
the privacy preserving aspect.

\section{Approach}
\label{sec:approach}

An effective middle ground between topological and cryptographic mechanisms
is still missing, e.g., a k-anonymous system providing strong privacy,
augmented with an anonymity set larger than k for cheaper and
more frequent attacks. In this section we introduce our approach to
solve this problem.

\subsection{Attacker Model}

In this section we restrict the protocol to the honest-but-curious 
model of attackers. Attackers from this model are following
the specification of the protocol and will not send maliciously created
messages, impersonate or create other identities, create fake messages or refuse
to respond. An attacker will try to extract as much information as
possible from
given messages to deanonymize the participants of the system
and try to attribute transmitted messages to their originator.

This model of attacker is fairly restrictive for attackers and it is
obvious from the descriptions of \Cref{sec:rec} that some of these
systems can easily be disturbed or even broken through malicious
messages. In \Cref{sec:discussion}, we will discuss an extension of the
attacker model and how a stronger attacker can be prevented or
identified before compromising the privacy of the users.

\subsection{Basic Protocol}

\begin{figure}[htbp]
\vspace{1.5em}
\centering
\def\svgwidth{.9\columnwidth}
\includegraphics[width=.8\columnwidth]{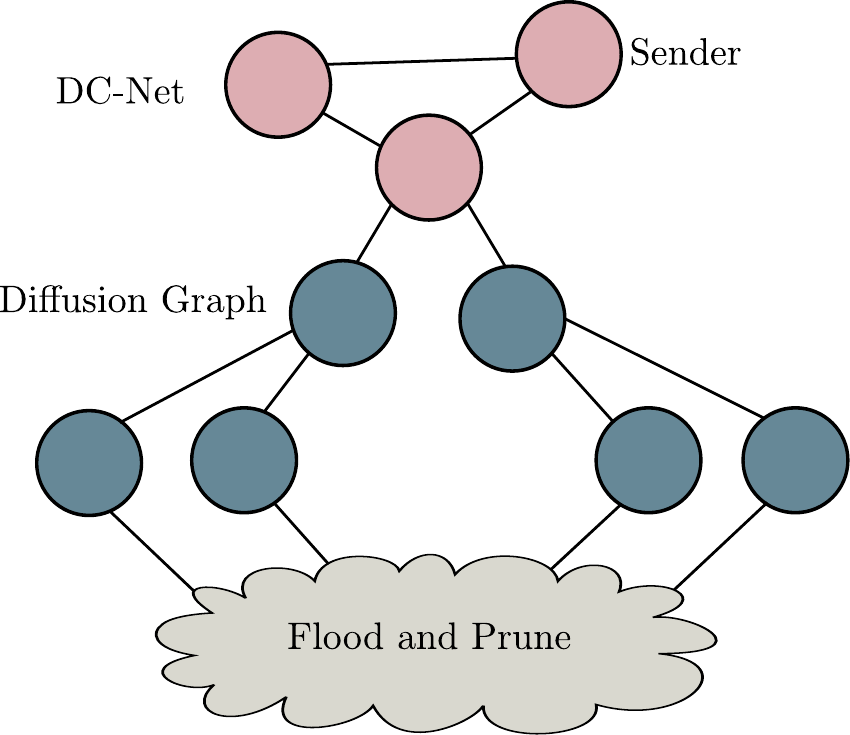}
\caption{Three phases of our privacy-preserving broadcast,
consisting of a DC-network using \(k=3\) (light red) and a diffusion tree
with \(d=2\) (dark blue).}
\label{fig:phases}
\end{figure}

Our proposal for a flexible privacy protocol, with a lower
bound on privacy, consists of the following three phases,
which are also illustrated in \Cref{fig:phases}.

\begin{enumerate}
 \item Spread message within a DC-network of size \(k\) (cf.
  the algorithm provided in Fig.~\ref{alg:dc}).
 \item Determine the first virtual source within the DC-network
  and continue with Adaptive Diffusion for \(d\) rounds.
 \item Perform a flood and prune broadcast until every participant in
     the network is reached.
\end{enumerate}

To complete the construction, the transitions between phases need to be defined.
The transition from the DC-net of Phase 1 to a single virtual source for phase
two needs to be independent of the originator to preserve its
anonymity.
Further, there should be no message overhead and the transition should
be verifiable for all group members to detect misbehavior.
To achieve this,
the node whose hashed identity, e.g., public key, is closest to the hash of the message
creates the initial virtual source token and starts the adaptive diffusion by
balancing the graph around them.
This fulfils the stated requirements as no additional messages need to be
transmitted. The mechanism depends only on the message and is
independent of the originator. Further, the transition can
be verified by all nodes of the group.

The parameter \(d\) of the adaptive diffusion phase is chosen based on the
network diameter to reach a large amount of nodes. As the round counter is carried
along the path of virtual sources, the final virtual source detects the last
round and sends a final of the spreading request. This message
does not only instruct the leaf nodes to spread the message, but also to switch
to flood and prune spreading.

This leads to the realization that adaptive diffusion messages can
be distinguished from flood and prune messages and nodes can detect
the current phase of the spreading protocol
even if they are not the virtual source. In \Cref{sec:discussion}
we show that this is not a problem for the privacy properties
of the protocol.

\subsection{Group Join and Leave}

The first phase of the proposed protocol consists of a form of
group communication. This requires a join or create group
operation, to form the groups. Group members need to react to
nodes leaving the group, such that the intended
group size remains within chosen parameters, namely \(k\) and \(2k-1\)
as a group of size \(2k\) can be split in two groups of size \(k\).
Until the network is large enough to satisfy the minimal
group size \(k\), privacy can not be guaranteed.

An approach to reduce spread in group sizes, which range from
\(k\) to \(2k-1\), is to allow multiple overlapping groups.
But group creation needs to be adjusted, as overlapping
groups can impact statistical privacy. As an example imagine
a group of size 3 with members A, B and C. Nodes B and C
are part of two groups, while A is only part of one group.
If nodes select the group to send randomly, a message from
this group of three has a probability of \(\frac{1}{2}\)
to have A as the origin of the message instead of the
desired probability of \(\frac{1}{3}\). A solution is
to enforce a number of groups to smooth probabilities.

A well designed join operation can improve the privacy of
participants by allowing them to select known or trustworthy
nodes. This improves privacy by preventing the subversion of
groups by controlling multiple nodes in a DC-network: The
node can select known distinct partners or participants
they know personally, increasing their trust in their security.
This concept is used by Herd~\cite{leblond2015herd} in the form
of anonymity providers.

On the other hand, this leads to an adverse effect: It is
important not to delegate the full creation of the group to
a single node, as this might leave the group under the control
of colluding nodes, stripping the node of the privacy benefits.

As a finishing note, group creation needs a more thorough
investigation, but a first solution would be the protocol 
by Reiter~\cite{reiter1996secure}. Reiter's protocol implements
a manager-based system tolerating up to one third of malicious
nodes using a consensus protocol.

\section{Discussion}
\label{sec:discussion}

Within this section we examine the desired properties of the
proposed protocol: performance and privacy.

\subsection{Performance}

To get an overview of the performance it is useful
to assess the phases separately from each other. The
transitions hardly create any overhead in messages
and latency, as the first transition consists only of computing
a hash and the second transition consists of an operation that is 
necessary in the protocol of Phase 2.

The first phase incurs \(\landauO (k^2)\) messages periodically. 
Our approach to reduce the overhead of the first phase, 
especially if there is no message to send, the base
message size could be restricted to an integer representing the length of the next message, e.g. 32 bit.
If the shared integer is not zero, a follow up round uses the resulting
number as a one time message size. To protect the length distribution from collisions, the integer
needs to be protected by CRC bits or similar mechanisms.

In a first simulation to estimate the performance of the
second phase, we averaged 12,500 messages with adaptive diffusion
to reach all 1,000 peers. This compares to an average of
7,000 messages for a regular flood and prune broadcast. As adaptive
diffusion will not be used to reach all nodes, just to
ensure privacy until a large segment of the network is reached,
this overhead can be assumed to be lower for the protocol.

Lastly, the intervals between rounds should be chosen
suitably for the expected activity in the network to minimize collisions,
but they can be adapted to changing activity.

\subsection{Privacy Properties}

The privacy guarantees from our building blocks~\cite{vonahn2003kanonmessages,Fanti2015adaptivediffusion}
translate to the following guarantees for our protocol: After Phase 1, if a group has 
\(\ell\leq k\) honest members, the protocol provides sender 
\(\ell\)-anonymity~\cite{vonahn2003kanonmessages}. For Phase 2, even with additional
information, the original group can only be recovered with low probability.
For suitable graphs and well chosen parameters the probability to
detect the true origin is close to the goal of perfect obfuscation~\cite{Fanti2015adaptivediffusion},
i.e. the probability of origin is \(\frac{1}{n}\).

The privacy assessment of the protocol can be split into two parts:
Assessment of the phases and assessment of the phase transitions.
The privacy evaluations of the original publications still hold for
Phase 1~\cite{vonahn2003kanonmessages} and Phase 2~\cite{Fanti2015adaptivediffusion}.
Phase 3 has no notable privacy properties. Due to these
pre-existing arguments from the literature we only argue for
the privacy of phase transitions.

For the transition from Phase 1 to Phase 2, we examine
the information used to perform the transition. The decision is
based on the shared message, which retains the anonymity guarantee
of Phase 1. The decision for the first virtual source
relies only on information under this guarantee, it can not introduce
additional information. Therefore, the transition to Phase 2
does not reduce the privacy below the privacy guarantees of Phase 1.

The transition from the second to the third phase is done by the last
node in possession of the virtual source token. This node has
no information in addition to the information inherent in the
adaptive diffusion protocol. The end of an adaptive diffusion
could be detected by any node through the lack of additional messages.
Therefore, the specific end message does not introduce additional
information leaks compared to the original protocol~\cite{Fanti2015adaptivediffusion}.

\subsection{Stronger Attacker Models}

In \Cref{sec:approach}, we restricted attackers to an honest-but-curious model,
restricting them to following the protocol. However, the presented phases are
vulnerable to different, stronger attacks.
Especially the presented implementation of DC-nets is vulnerable to
denial of service, by creating collisions through sending random messages,
without much chance to identify misbehaving participants.

Such an attack on the liveliness can be countered by a blame protocol. For restricted
DC-nets, von Ahn et al.~\cite{vonahn2003kanonmessages} provide a solution
using additional restricted commitments and possible blames. The
proposed procedure can be used to either remove the faulty entity
from the group or dissolve the group. This also depends on the design
of the group creation and join mechanisms. 

The solution creates additional message overhead though.
As the network operates in the context of blockchains,
an honest but curious attacker might provide a better model:
Even without a blame protocol nodes might just dissolve the group
and create a new one without nodes that they do not
consider trustworthy. This is enough to protect their
privacy but might result in missed transactions and
transaction fees for the attacker, while not providing additional information.
As a result of this consideration, it should be considered which
attacker model is suitable and if use case specific methods can improve
the overall efficiency of the protocol, for a general use case,
the blame option should be the default.

This decision has consequences for the evaluation of the rest of the protocol.
It is easily detectable if an owner of the virtual source token
refuses to disseminate the message. The network can react to those
disturbances. So for an attack on the privacy of users, it is
more useful to stay undetected, hence the restriction on the
original honest-but curious-model.

\section{Conclusion}
\label{sec:conclusion}
In this article we reviewed several mechanisms to provide privacy for transactions in blockchains
on the network level. While these mechanisms provide good results for their respective use cases,
they are not flexible and only provide solutions for the edges of the
privacy--efficiency trade-off spectrum.

Based on these mechanisms we proposed a new adjustable privacy-preserving
broadcast protocol with a lower bound on privacy and improved protection
against statistical analysis using well connected attackers, e.g., by
deploying botnets.

For a full evaluation of the covered privacy-efficiency space
further research is required. A full privacy analysis and a performance
analysis with corresponding implementation will provide data for
application designers to choose suitable and safe parameters to preserve
the privacy of users. Lastly, further research for secure group creation
might lead to improved privacy guarantees or more realistic trust assumptions,
and additional optimizations would lead to a more efficient protocol.

\section*{Acknowledgment}

This work was partially funded by the Baden-W\"urttemberg Stiftung.

\bibliographystyle{IEEEtran}
\bibliography{network}

\end{document}